\documentclass{elsart}
\usepackage{amsmath,amsthm}
\usepackage{makeidx,graphics}
\usepackage{epsfig}

\newtheorem{theo}{Theorem}

\def\be{\begin{eqnarray*}}
\def\ee{\end{eqnarray*}}

\newcommand{\refe}[1]{equation~(\ref{#1})}

\newcommand{\beq}{\begin{eqnarray}}
\newcommand{\eeq}{\end{eqnarray}}
\newcommand{\expect}[1]{\left \langle #1 \right \rangle }
\def\half{ \frac {1}{2} }

\begin{document}
\runauthor{Coppersmith, Kadanoff and Zhang}
\begin{frontmatter}
\title{Reversible Boolean Networks II: Phase Transitions, Oscillations, and Local Structures}
\author[Chicago]{S. N. Coppersmith, Leo P. Kadanoff, Zhitong Zhang}
\address[Chicago]{The James Frank Institute, the University of Chicago,
5640 S. Ellis Ave, Chicago, IL 60637}

\begin{abstract}

We continue our consideration of
a class of models describing the reversible dynamics of
$N$ Boolean variables, each with $K$ inputs.
We investigate in detail the behavior of the Hamming distance
as well as of the distribution
of orbit lengths as $N$ and $K$ are varied.
We present numerical evidence for a phase transition in the
behavior of the Hamming distance at a critical value
$K_c\approx 1.65$ and also
an analytic theory that yields
the exact bounds on $1.5 \le K_c \le 2.$
%as well as an estimate for $K_c$
%that agrees reasonably well with the numerical observations.

We also discuss the large
oscillations that we observe in the Hamming distance
for $K<K_c$ as a function of time
as well as in the distribution of cycle lengths
as a function of cycle length
for moderate $K$ both greater than and less than $K_c$.
We propose that local structures, or subsets of spins whose
dynamics are not fully coupled to the other spins in
the system, play a crucial role in generating these oscillations.
The simplest of these structures are linear chains, called linkages,
and rings, called circuits.
We discuss the properties
of the linkages in some detail, and sketch the properties of circuits.
We argue that the observed oscillation phenomena can be
largely understood in terms of these local structures.
\end{abstract}
\begin{keyword}
Gene Regulatory networks; Random boolean networks; Time-reversible Boolean
networks; Cellular automata; 
\end{keyword}
\end{frontmatter}
\newpage

\section{Introduction}

Kauffman nets~\cite{Kau1} have been used to model the complex behavior of
dynamical systems ranging from gene regulatory
systems~\cite{Kau93,Kau84,Kau95}, spin glasses~\cite{Der86,Der87}, 
evolution~\cite{other NK}, social sciences~\cite{Wie1,Gar1,Hor1,Kha1}, to the stock market~\cite{Stuf}.
A Kauffman net consists
of $N$ boolean variables or spins, each of which
is either ``+1'' or ``-1'' at each time $t=0,1,2,\ldots$.
We define $\sigma_i^t$ to be the
value of the $i^{th}$ spin at time $t$.
The system evolves through a sequence of states
$\Sigma^0,\Sigma^1,\ldots,$ where
\begin{equation}
\Sigma^t=(\sigma^t_1, \sigma^t_2, \ldots, \sigma^t_N)~.
\end{equation}
There are $2^N$ different possible $\Sigma_t$'s.

The Kauffman net evolves according to
\begin{equation}
\sigma^t_i=F_i(\Sigma^{t-1}) ~,%_{P^i_1}, \sigma^{t-1}_{P^i_2}, \cdots,
%\sigma^{t-1}_{P^i_K})~,
\label{kaueq1}
\end{equation}
where $F_i$ is a boolean function with $K$ boolean arguments picked from
among the $N$ possible arguments in ${\Sigma^{t-1}}$ and one boolean
return.
The $F_i$  for $i=1, 2,
\ldots, N$ are fixed during the evolution.
The combination of them is called a
realization. There are $(2^{(2^K)})^N\binom{K}{N}^N$ different
realizations for a net of $N$ elements with $K$ inputs~\cite{Kau93}.

After specifying a realization, one picks a
starting state for the system; the network
then evolves following
\refe{kaueq1}.

Reference~\cite{Sue00} (which we shall call paper I) introduces a
time-reversible~\cite{Ben1,Min1,Mar1} boolean network model,~\cite{margolus} with dynamics
governed by the equation
\begin{equation}
\sigma^{t+1}_i=F_i(\Sigma^t) \sigma^{t-1}_i~.
% {P^i_1},\sigma^t_{P^i_2}, \ldots,\sigma^t_{P^i_K}) .
\label{treq}
\end{equation}
Each time-reversible network realization has a corresponding
dissipative realization with
same functions and connections.
In the reversible model substates
$\Sigma^{t-1}$ and ${\Sigma^{t}}$ are both required to calculate
$\Sigma^{t+1}$~\cite{TOF90}, so the state of the system
is $({\Sigma^{t-1}},{\Sigma^t})$, and the state space has
$2^{2N}$ points.

Paper I studies how the distribution
of cycle lengths generated by the time-reversible model
scales with $N$ for different values of $K$.
% The general conclusion is that for small $K$,
%ay $0$ or $1$, the median or typical cycle length is independent of $N$,
%hile for large $K$ typical cycle lengths grow exponentially with $N$.
%The analysis makes heavy use of the distinction between two kinds of cycles:
%special cycles, which are invariant under time-reversal,
%and regular cycles, which are not.
%
This paper investigates the cycle length distribution
in more detail and also investigates the behavior of
the Hamming distance~\cite{Der86,Ham2,Wae1} in this model.
In the next section, we
present both numerical evidence and analytic
arguments that the behavior of the Hamming distance
changes qualitatively and
a phase transition occurs in the model
at a critical value $K_c \simeq 1.65$.
In section~\ref{sec:Oscillations}
we investigate how the realization average of the average number
of cycles depends on the cycle length.
For small $K$,
cycle lengths divisible by integer powers of two and by three
are much more
likely than lengths that are products of larger prime numbers.
Oscillations are also observed in the behavior of the
Hamming distance as a function of time when $K$ is small.
We explain these observations in terms of ``local structures,''
groups of spins whose dynamics do not depend on the state
of any other spin in the system.
In section ~\ref{sec:k=1}, we discuss the two most important
structures, \emph{linkages} and \emph{circuits}.
We calculate Hamming
distances and the cycle-length distribution for linkages.  Our
calculations for these structures are then connected back to the
numerical observations, both for $K=1$ and for higher $K$.

\section{Hamming distance in reversible Boolean nets.}
\label{sec:hamming}
In this paper we define the Hamming distance $D(t)$ to be
the number of spins in two substates
that are different at time $t$.
%Since a state of the system $S_t$ consists of
%the two spin configurations
%$\Sigma_{t-1}$ and $\Sigma_{t}$,

Consider two different time developments of a given realization
of a system that are identical at
$t=0$ and that have a single spin different at $t=1$.
We obtain two series of
configurations, called two \makebox{\em{paths}} of
the system, and calculate $D(t)$, the Hamming distance
between the two paths as a function of time $t$.
In this subsection we consider the behavior of $\expect{D(t)}$, the
Hamming distance averaged over different initial configurations and
realizations, as a function of time $t$,
for different values of the parameters $K$ and $N$ in our
time-reversible model.
We will find that
when $K$ is  small, at long times the maximum value of the Hamming
distance saturates at a value that is independent of $N$
as $N \rightarrow \infty$, while
when $K$ is sufficiently large, the saturation value
is proportional to $N$ as $N \rightarrow \infty$.
These two behaviors are separated by a phase
transition at $K_c$, a critical value of $K$.

It is useful to define
the normalized Hamming distance
$\expect{d(t)}=\expect{D(t)}/N$, the average
over realizations of the fraction of variables having different
values in the two paths at time $t$.
Below the phase transition, one finds that as $N\to \infty$ 
$\expect{d(t)}$ tends to zero for all $t$.
while above $K_c$ the maximum of $\expect{d(t)}$ has a non-zero limit.

\subsection{Hamming distance for $K=0$ and $K=N$}
When $K=0$,
if one starts with two copies of the system that are identical at
time $t=0$ and have exactly
one spin different at time $t=1$,
the Hamming distance oscillates
between $0$ and $1$ at all future times.
When $K=N$,
flipping a single spin changes an input of every spin,
so that all spins have probability $1/2$ of being the same in the
two paths at time $t=2$.
Thus, when $K=N$, for all times $t\ge 2$, $\expect{D(t)}=N/2$, and
$\expect{d(t)}=1/2$.
Thus we see that the Hamming distance at long times
$\expect{D_\infty}\equiv \lim_{t\rightarrow \infty}D(t)$,
is independent of system size when $K=0$ and is
proportional
to the system size $N$ when $K$ is very large.

\subsection{Hamming distance for intermediate $K$}
\label{hamming:numerical}
\begin{figure}
\centerline{\epsfxsize=3.5in
\epsfbox{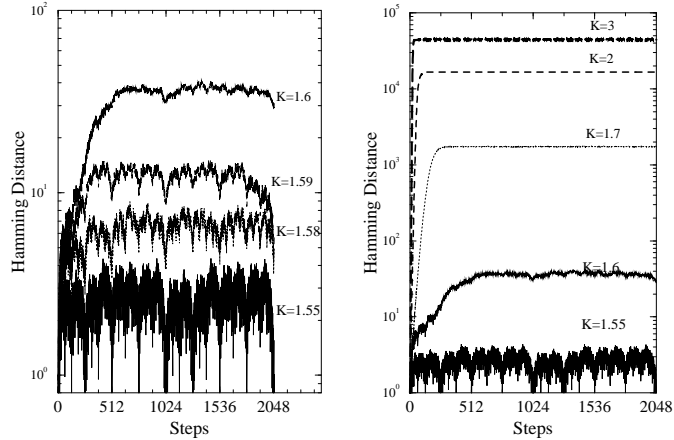}}
\caption{Plot of $\expect{D(t)}$, the average over realizations
of the Hamming distance as a function of time $t$,
for systems with $N=100,000$
and different values of $K$.
The starting initial conditions for each realization were chosen
so that $D(0)=0$, $D(1)=1$. 
For visual clarity, this figure is split into
two parts with different vertical scales.
The left subfigure includes only $K$-values
below or close to the critical value
$K_c \simeq 1.65$, while the right subfigure
includes $K$-values significantly above $K_c$.}
\vskip 0.75cm
\label{trfig1}
\end{figure}

To investigate intermediate $K$, we simulated
reversible networks of 100, 1,000, 10,000 and 100,000 boolean
variables for different $K$ values. For each case, 100 realizations
were generated, and for each realization, one pair of paths
with $D(0)=0$ and $D(1)=1$ was examined. Numerical
results for $N=100,000$ are presented in figure~\ref{trfig1}.  For all
$K$'s, the Hamming distance first increases, then reaches a plateau,
and finally oscillates about that plateau.
For small $K$, large oscillations in
$\expect{D(t)}$ are observed.  In particular, when $K$ is small,
$\expect{D(t)}$ is especially small when the time $t$ is a multiple of
$2^m$ for integer
$m$ (e.g., at $t=64$, $128$, and $256$).
 For larger values of $K$ these oscillations
become much less pronounced.

\begin{figure}
\centerline{\epsfxsize=3.5in
\epsfbox{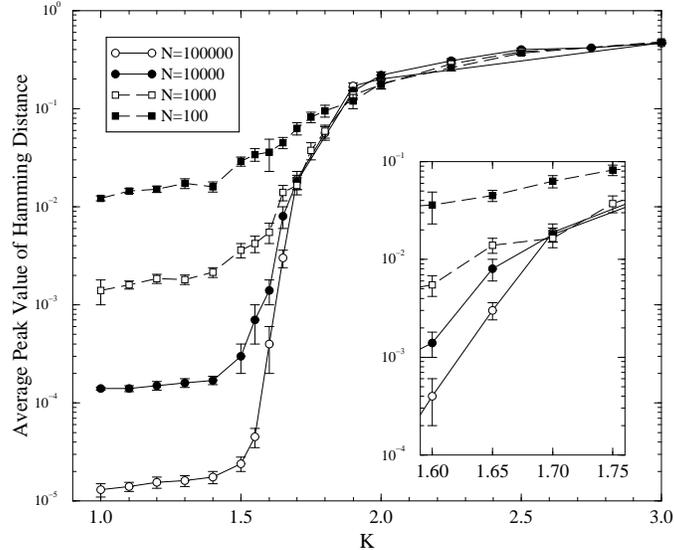}}
\caption{Average of the peak of the normalized Hamming distance
$\expect{d}_{\max} \equiv \expect{D}_{\max} / N$
as a function of $K$ for four decades of $N$
values. For each set of $N$ and $K$, we generated 100
pairs of paths with initial separations $D(0)=0$ and $D(1)=1$. Each of them is derived from a different realization
and initial condition. For each pair of paths we
calculate $D_{\max}$, the peak value of the Hamming distance, and then
average this over realizations.
The inset shows the behavior near criticality. We use these
data to estimate that the critical value $K_c$ is about 1.65.}
\label{trfig2}
\vskip 0.75cm
\end{figure}

This figure shows that $\expect{D}_{\max}$, the
maximum value of the Hamming distance, exhibits a sharp jump
as a function of $K$ at $K_{c}\approx 1.65$.
This jump arises because, for $K>K_c$, $\expect{D}_{\max}$ is proportional
to $N$ for large $N$, while for $K<K_c$,
$\expect{D}_{\max}$ is independent of $N$ as $N \rightarrow \infty$.
This
behavior is shown  in figure \ref{trfig2}, which plots the maximum
level of the normalized Hamming distance
$\expect{d}_{\max} \equiv \expect{D}_{\max}/N$
as a function of $K$ for different $N$.
Thus, $d_{\max}$ appears to act
as an order parameter for a phase transition at $K_c\approx1.65$.

The phase transition that we observe at
$K_c \simeq 1.65$ is of a percolative nature~\cite{Perc,Ma76,Sah1,Bun1}.
When $K>K_c$, if one changes a few spins in the system,
then information about the change spreads to more and more
spins and eventually covers a nonzero fraction of all the spins
in the system.  When $K<K_c$,
information remains localized within a few spins which tend to
oscillate with periods which are powers of small prime integers.
This phase transition is then connected with the percolation of
information within the system.

The percolative nature of the phase transition in the reversible model 
has similarities to that observed in the dissipative Kauffman 
model~\cite{Der86,DS86,BP98}, but there are differences.
Specifically, consider two paths that are the time histories of
a system realization that is started with two initial conditions
that are almost the same, but having a small number of spin-values
differing or `deviating' from one another.
The original Kauffman model is dissipative, so
a spin whose values are deviating at early
stages of the history may stop deviating as time goes on.
However, a reversible system must undergo cyclic behavior.
If a spin deviates at
one time, it will do so at an infinite number of later times.
In a reversible system, deviation is a permanent condition.

So we must ask how often a deviating spin triggers the
deviation of others.
Let $G$ be the average number of additional spins whose deviations are
directly caused by a lone deviation in the system.  The value of $G$
determines whether a few deviations will cause a chain reaction which will
spread to the entire system.  If $G>1$, the chain reaction will occur and
the Hamming distance will eventually
saturate at a value proportional to $N$.
In the opposite case, $G<1$,
a single deviation triggers less than one additional deviation
on average, and the
effect saturates long before the entire system is
affected.

%\begin{table}
%\begin{center}
%\begin{tabular}{|l|l||c|c|} \hline
%function & effective & \multicolumn{2}{c|}{values  of spin $\sigma$}   \\
%\cline{3-4}
%name & $K$-value & $\sigma=$+1$$ & $\sigma=-1$  \\ \hline \hline
%%\begin{tabular}{|l|l||l|l|} \hline
%one   & $0$  & $+1$  &  $+1$  \\
%-one   & $0$  & $-1$  &  $-1$  \\
%$\sigma$   & $1$  & $+1$  &  $-1$  \\
%$-\sigma$   & $1$  & $-1$  &  $+1$    \\  \hline
%\end{tabular}
%\end{center}
%\label{keq1}
%\caption{NEED TO FIX THIS TABLE!
%The different functions, $F(\sigma)$, for $K=1$. The numbers
%$\pm1$ show the different value of the functions for different values of
%their arguments. Under the symmetry operation in which
%$\sigma$ is replaced by -$\sigma$, they fall into two classes, depending upon
%their effective $K$-value.}
%\end{table}

\begin{table}
\begin{center}
\begin{tabular}{|l|l|l||c|c|c|c|} \hline
function & number of inputs & number &
 \multicolumn{4}{c|} {output for the input values
}  \\  \cline{4-7}
 example & that affect output & in class & ($+1$,$+1$) & ($+1$,$-1$) &($-1$,$+1$) &($-1$,$-1$)
\\ \hline
\hline
+1   & $0$ & $2$ & $+1$  &  $+1$  & $+1$ & $+1$ \\
$\sigma_1$   & $1$ &4  & $+1$  &  $+1$ & $-1$ & $-1$  \\
$\sigma_1 \sigma_2$   & 2 &2  &$+1$ & $-1$  & $-1$& $+1$    \\
$  \sigma_1 ~~\text{or} ~~ \sigma_2$  & $q$ &8  &$+1$ & $+1$  & $+1$& $-1$   \\
\hline
\end{tabular}
\end{center}
\caption{
The different classes of functions, $F(\sigma_1,\sigma_2)$, for $K=2$.
%The numbers
%$\pm1$ show the values of the outputs of the
%functions for different values of
%their arguments.
The classes are defined by the symmetry operations of spin
flip
and interchange of the two spins.}
\label{keq2}
\vskip 0.75cm
\end{table}

\subsection{Bounds on the critical value $K_c$.}
Though we have not calculated exactly the branching ratio $G$
for this chain reaction, we can prove
that $1.5<K_c<2$.
We first prove that $1<K_c\le2$, and then refine the argument to
show that $1.5<K_c< 2$.

We define systems with fractional $K$ to be composed of elements
that have probability $[K]+1-K$ of having $[K]$ inputs and
probability $K-[K]$ of having $[K]+1$ inputs, where
$[x]$ is the largest integer less than or equal to $x$.
(This then defines $K$ as the average number of inputs, since
$([K]+1-K)+(K-[K])=1$ and $[K]([K]+1-K)+([K]+1)(K-[K])=K$).
The growth rate $G_K$ is the
average
\begin{equation}
G_K = ([K]+1-K)G_{[K]} + (K-[K])G_{[K]+1}~.
\label{e:varyK}
\end{equation}
Thus we need only calculate $G_K$ for integer $K$.

\subsubsection{$K_c>1$.}
We can see immediately that deviations
will not percolate if $K$ is less than or equal to one.
When $K=0$, every spin is independent of its inputs, so
$G_{K=0}$=0.
When $K=1$, there are
four different possibilities for $F_i(\sigma)$, namely $1$, $-1$,
$\sigma$, and $-\sigma$.
In our model all functions occur with equal probability,
so a deviation in the spin $\sigma$ triggers a deviation in
the spin $\sigma_i$ to which it is an input with probability $1/2$.
Thus the growth factor for $K=1$ is
\begin{equation}
G_{K=1} = 0.5~.
\label{e:keq1}
\end{equation}
The growth factor for $K$ between $0$ and $1$ is just $KG_{K=1}$, so
for $K$ less than or equal to one the system must be below its
percolation threshold.

\subsubsection{$K_c < 2$.}
For any $K$, changing any one of the $K$ arguments of a function
$F_i({\Sigma})$ has a $50 \%$ chance of flipping its output value.
Therefore, a deviation of a single spin will, at the very next step,
cause an average of $K/2$ deviations.  Later steps may cause further
deviations, but considering only this initial step
certainly bounds the growth factor:
\begin{equation}
G_{K} \ge K/2~.
\label{e:ggtK}
\end{equation}
Thus, when $K>2$ the growth factor $G_{K>2}>1$, so that
the deviation will spread to a nonzero fraction of the entire system.

At first sight, it appears that $K=2$ is the marginal, or critical
case. That is wrong.  Information will also percolate through the
entire system when $K=2$.
In the previous paragraph, we considered the
further deviations induced in the very next step after the
deviation of a given spin.
These effects were enough to ensure that at least
$K=2$ is marginal.  However, a deviation can have a delayed further
effect, producing additional deviations beyond the one we have
already calculated. These additional deviations then permit
information to percolate through the entire system for $K=2$.

To see this result, we need only establish one case with a further,
delayed, deviation.
Notice Table~\ref{keq2}.
Consider a situation in which spin number
$3$ is determined by the ``or" function shown on the last line in the table.
Assume that input spin $1$
has just deviated and input spin $2$ has the value $+1$ on both
paths. Now spin $3$ does not deviate at this step.  
However, if at a later stage spin $2$ takes on the value $-1$, that will
induce a delayed deviation of spin $3$.
Since situations
like these occur with non-zero probability,
and increase the value of $G_{K=2}$ beyond
the value shown in equation \ref{e:ggtK}, the case
$K=2$ is above the percolation threshold.

\subsubsection{$K_c>1.5$.}
Since we know $1<K_c\le2$, we need only consider
values of $K$ between $1$ and $2$.
We have already calculated $G_{K=1}$ and a lower bound to
$G_{K=2}$; now we wish
to consider $G_{K=2}$ in more detail.

Let us estimate the average number of additional deviations produced by
one deviated spin.
The growth rate $G_{K=2}$ is $2N$ times
the probability that a spin $\sigma$ is induced to deviate when one of
its two inputs has deviated, which is in turn
the sum of
products of the likelihood of a function choice times the probability
that the function gives different outputs with one deviated input.
Of the sixteen two-input functions $F(\sigma_1,\sigma_2)$,
two functions ($\pm1$) depend on zero inputs,
four functions ($\pm\sigma_1$,
$\pm\sigma_2$) depend on one input (they
change output if one input is changed but not the
other one), and two functions
($\pm\sigma_1\sigma_2$) depend on two inputs (they
change if either input is changed).
The remaining eight functions
are like the ``or" function; they depend
upon one argument only for a particular value of the other argument.
(See Table~\ref{keq2}.)
Therefore,
$$
G_{K=2}  = \frac{2 \times 0+4 \times 1 +2 \times 2 + 8 \times q}{16}~,
$$
where $q$ is the unknown effective number of inputs of the
functions like ``or".
We overestimate $q$ by saying that these functions
all behave exactly as if they had two inputs.
This estimate then gives us
\begin{equation}
G_{K=2}  \le  \frac{3}{2}~.
\label{e:gkeq2}
\end{equation}
Putting together our results from equations \ref{e:keq1}, \ref{e:varyK}
and \ref{e:gkeq2}, we find that
\begin{equation}
G \le (2-K) 0.5 + (K-1) 1.5 =-0.5 + K~.
\label{e:varyK1}
\end{equation}
This overestimate gives us a marginal growth
rate at $K=1.5$, which yields a lower bound
to the critical value $K_c\ge1.5$.
Our best numerical
evidence is computed from the data shown in figure
\ref{trfig2} and gives
$K_c= 1.65 \pm 0.10$. 
This result is, of course, in agreement with the exact
bounds.

\section{Oscillations in orbit period distributions}
% for intermediate $K$ values}
\label{sec:Oscillations}
In this section we present our numerical results for cycle length
distributions, and give a qualitative explanation of what we are seeing. 
When $K$ is small but larger than $K_c$, the number of cycles does not
depend smoothly on length. Instead, cycles whose lengths are divisible by
small integer powers of two and by three predominate.  As $K$ is
increased, the cycle length distribution becomes smoother.  We relate this
trend to the evolution in the connectivity of the network as $K$ is
increased.  When $K$ is small, we can have small clusters of spins that
are only weakly coupled to the other spins in the system.  These 
\emph{local structures} are shown to have a characteristic effect of yielding
cycle lengths divisible by small factors such as two and three.  As $K$
increases, all the spins tend to be in a single large cluster with complex
interactions, and local structures become much less likely.  This
leads to a much smoother cycle length distribution.

\subsection{Simulation results for cycle length distributions}
\label{sec:cycle_lengths}

\begin{figure}
\centerline{\includegraphics[width=4.5in]{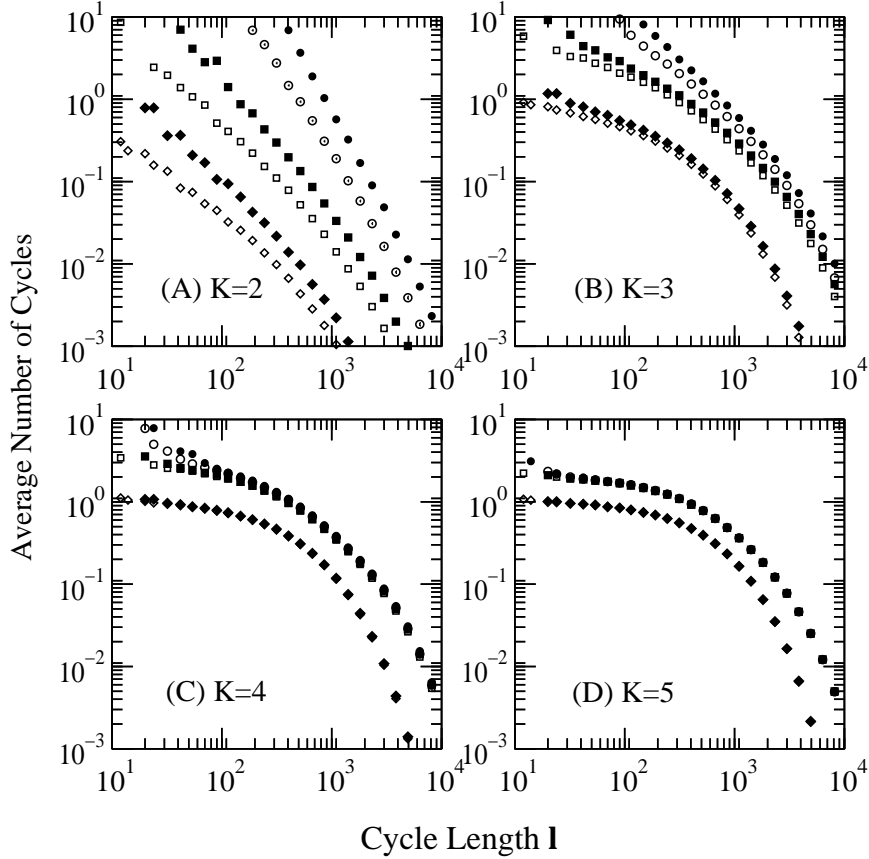}}
\caption{Log-log plots of numbers of
cycles per realization,
averaged over realizations, as
a function of cycle length $l$ for $N=10$ and for
$K=2$, $3$, $4$, and $5$.   
For each $K$, all cycles are enumerated from each of
25600 realizations.
Note the different branches arising from oscillations
in these functions.
The different point
styles distinguish the different divisors of the $l$-values.
They are:
$l=0\mbox{ mod }4$ (circles), $l=2\mbox{ mod }4$ (squares) and
$l=1\mbox{ mod }2$ (diamonds).
Filled symbols have $l=0\mbox{ mod } 3$, while
open symbols denote $l$ not divisible by three.}
\vskip 0.75cm
\label{fin41grf}
\end{figure}

Figure~\ref{fin41grf} shows the distribution of cycle lengths
 for intermediate
$K$ values and $N=10$.
The distribution oscillates strongly when $K$ is small.
As $K$ is increased, the oscillations diminish and
the distribution eventually converges to the
random hopping behavior discussed in Paper I~\cite{Sue00}.
A complex oscillation pattern in the distribution is evident, particularly
for the cases $K=2$ and $K=3$.
Several different branches are observed, which
reflect periods which
are or are not divisible by small powers of two and of three.
The oscillations at moderate $K$ in the
reversible model are significantly more pronounced than those
observed in dissipative Kauffman models~\cite{Liang}.
The
amplitude of the oscillation diminishes as $K$ increases, and
as $K$ approaches $N$, the distribution
approaches the two-branch structure (due to an even-odd
oscillation) calculated in the $K=N$ case in Paper I.

\begin{figure}
\centerline{\epsfxsize=3.5in
\epsfbox{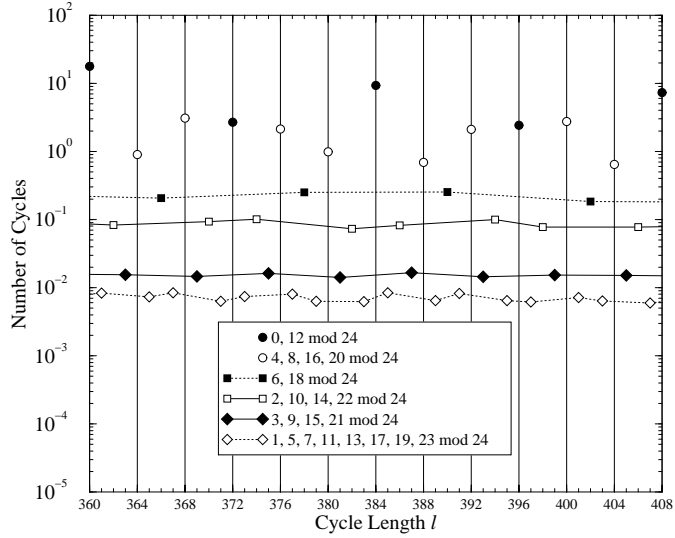}}
\caption{Enlarged plot on a semilog scale of the realization average
of the number of cycles of length
$l$ versus $l$ for $N=10$ and for $K=2$.
The oscillation structure is highlighted by showing only
the region extending from $l=360$ to $l=408$.
The main structure of the figure consists of three
branches, which in decreasing order of probability satisfy:
$l=0\mbox{ mod }4$ (circles), $l=2\mbox{ mod }4$ (squares) and
$l=1\mbox{ mod }2$ (diamonds).
Filled symbols are used for lengths evenly divisible by three and
open symbols for those that are not.
The plotting symbols are the same as in Fig.~\ref{fin41grf}.}
\label{fin61grf}
\vskip 0.75cm
\end{figure}

To see the oscillatory structure in more detail,
we show in Fig. \ref{fin61grf}
an enlarged plot which shows the number of cycles for the region extending
from $l=360$
to $l=408$.
The most likely cycles fall into three main
branches, which in decreasing order of probability satisfy
$l=0\mbox{ mod }4$, $l=2\mbox{ mod }4$, and $l=1\mbox{ mod }2$.
Each main branch is further
split, with cycle-lengths that are multiples of three
being more likely.
Thus, the cycle
lengths that are divisible by
$24$ occur with the highest likelihood.

This oscillation is not a small effect.  Figure~\ref{fin61grf}
demonstrates that cycle lengths  divisible by
$24$ are more than two orders of magnitude more common than any odd cycle
length in this range of $l$. Thus we see a marked tendency for orbit
periods to contain factors of $2^n$ for small integer $n$ and of $3$.

\subsection{Characterizing the oscillations in the cycle length distribution}

We now look for a quantitative measure of the amount of oscillation
in cycle length,
so that we can see how it changes as a function of $K$.  More specifically,
we wish to ask
whether we can understand how the oscillations go away for large $K$.

\begin{figure}
\centerline{\epsfxsize=3.5in
\epsfbox{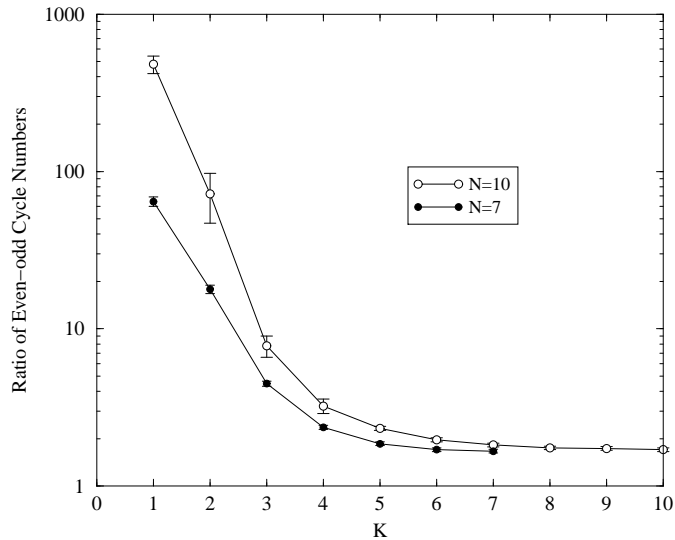}}
\caption{The ratio of the number of even cycles to that of odd ones
plotted against $K$ for $N=7$ and $N=10$. For each $K$ and system size,
all cycles are enumerated in each of $25,600$ realizations.
For
small $K$ values, the ratio increases rapidly as the system gets larger,
while for large $K$, the ratio stays at a level above unity, which
depends much more weakly on system size.}
\label{Osci:ratio}
\vskip 0.75cm
\end{figure}

For small $K$, the even cycles predominate.  It is very likely indeed,
that among the many independent localized stuctures in oscillation at least
one will give a
period two oscillation, which will then be observed in the overall period
of the system.  To
show this, we plot in Figure~\ref{Osci:ratio} the ratio of the 
number of even cycles to the number of odd ones for two different system 
sizes, $N=10$ and  $N=7$.  
For large $K$, the global behavior
forces the number of even cycles to be greater than the number of odd ones,
but the effect depends very weakly on $N$.
For smaller values of $K$ the ratio becomes much
larger, and has substantial $N$-dependence.
%This plot, then, supports our overall contention
%that local
%structures have a substantial effect at small but not large $K$.
 
To parametrize the oscillation structures for different $K$ values, we
first separate the domain of possible cycle lengths into segments of
length $P$. Based on the argument about the appearance of factors of 
small powers of $2$ and $3$ in the system's periods,
and also the observation in Fig~\ref{fin61grf}, we choose $P$ to be 24.
We also require the $i$th
segment to start at $24i$ and thus the 24 elements of the segment are
$24i+r,\quad r=0, 1, \ldots, 23$. Paper I discusses how correlated
fluctuations in the number of mirrors could produce an
even-odd oscillation in the number of cycles.  In this paper, the even-odd
oscillations (exhibited in figure 4 of Paper I) are just a
distraction. To eliminate the even-odd effect due to 
the correlated fluctuations in the number of mirrors, we define $q_i(r)$
to be the proportion of cycles of period $24i+r$ among all the even cycles
in the segment when $r$ is even, and among all the odd ones when $r$ is
odd. We then average $q_i(r)$ over segments of which the number of both
even cycles and odd cycles are observed more than 100 times in our
simulation (complete enumeration of 25,600 realizations), and thereby
obtain the functions $Q(r)$. 

\begin{figure}
\centerline{\epsfxsize=3.5in
\epsfbox{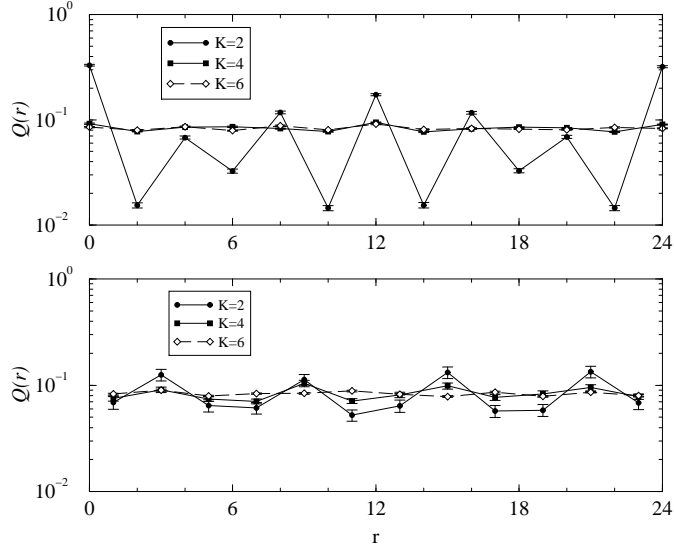}}
\caption{$Q(r)$ as a function of $r$.  Top: even cycles.
Bottom:  odd cycles.
High $Q$ values mean high likelihood of observing a
cycle length with certain residues. Numerical data were obtained by
complete enumeration of 25,600 realizations for each $K$. }
\label{osciQ}
\vskip 0.75cm
\end{figure}

The functions $Q(r)$ are plotted separately for even and odd values
of $r$ in
figure~\ref{osciQ}.
For $K=2$ we can see very strong oscillations as a function
of $r$ in both plots.  As expected, factors of three and of four
are favored (this plotting method does not show the oscillations
with period two).
As $K$ is increased, the $r$-dependence becomes much weaker.

To characterize the strength of the oscillation for a given
$K$ using a single parameter,
we order $r=0, 1, 2, \cdots, P-1$  in the increasing order with
respect to $Q(r)$ and index them as $r_1, r_2, ..., r_P$.
The Gini coefficient~\cite{Creedy98},
a measure of the inequality between the $P$ values of $Q(r)$, is defined as
\begin{equation}
G=1-\frac{2\sum^{P-1}_{i=1}\sum^{i}_{j=1}Q(r_j)}{(P-1)\sum^P_{j=1}Q(r_j)}.
\end{equation}
When all the $Q(r)$ are equal,
$G$ takes on its minimum value, $G=0$;
when only $Q(r_P)$ is nonzero, so that the distribution of
$Q(r)$ is extremely inequitable, then
$G=1$.

\begin{figure}
\centerline{\epsfxsize=3.5in
\epsfbox{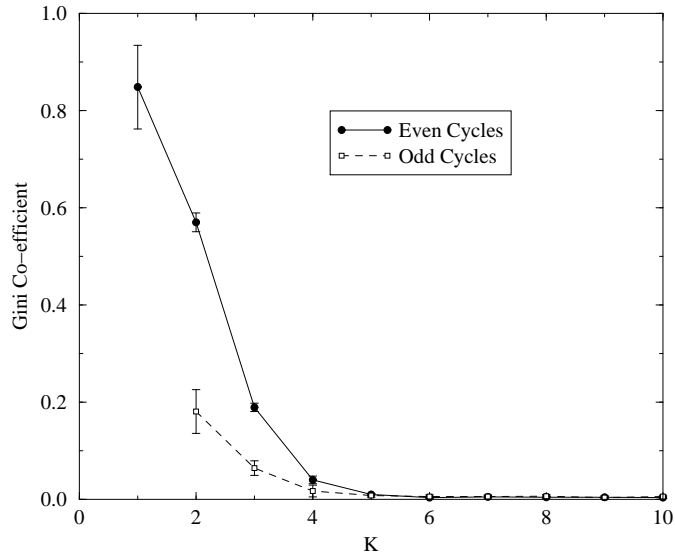}}
\caption{Our Gini coefficients measure the size of oscillations in the
number of cycles as a function of length. They are obtained by complete
enumeration of 25,600 realizations for each $K$ and plotted here, 
with even and odd cycles separated. When $K=1$, the
only odd cycles observed are of length $3$ and thus $G=1$ for the odd cycles.}
\label{gini}
\vskip 0.75cm
\end{figure}

The Gini coefficients as a function of $K$ with $N=10$ and
$P=24$ are shown in figure~\ref{gini}. Notice that there are separate
curves for even and odd values of $r$.   We see that the Gini
coefficient is large for small $K$ and decreases rapidly
as $K$ gets larger, showing values indistinguishable from zero for $K$
above $5$.
%These curves, then, support our view that the  structures
%which produce oscillations have less effect for higher $K$-values.

\subsection{Local structures}
In this section, we discuss why cycle lengths are so likely to
be evenly divisible by small powers of two and by three.
When $K>K_c$, the influence of a single spin flip percolates and
reaches a finite fraction of the entire spins in the lattice.
We can then imagine that a nonzero fraction of all the
spins in the system
fall into a main \mbox{\em connected} cluster, one with many complex
linkages among the
different spins in the cluster.  We depict this in Figure
\ref{clusters}, in which the
`C' denotes the connected cluster.  It is further indicated by the
surrounding curve.

\subsubsection{Isolated structures}
But $K$ is not infinite.  There will be spins in the system which are
only weakly
coupled in to the rest.  For example, there will be spins 
which are simply uncoupled to any others. One
such spin is indicated by the `$I_0$' in the figure.
This spin has a function
$F_i(\Sigma)$ which is simply $\pm 1$, independent of all the spins.
Thus, this
coupling has an effective
$K$-value of zero.
It is possible that this isolated
spin does not affect any other spins in the system.
It can also happen that the isolated spin
can affect one or several spins in the cluster.
We show this with the situation
labeled as `$I_1$'.  The effect of the two situations is rather similar.

There are other examples of weakly coupled spins.  But the effect of the
isolated spin can be
calculated in full detail, and will serve to illustrate the qualitative
nature of
localized-structure-effects.

First of all notice that for large $N$ there will be many such isolated
spins.  For a given
value of $K$ the probability that a spin is assigned the function
$-1$ or $+1$ is $2^{(1-2^K)}$, so that the
average number of such spins will be
$$
p=2^{(1-2^K)} N ~.
$$
This number must be large for sufficiently
large $N$.  A spin with input function $-1$ is in a period-4 cycle,
while a spin with input function $+1$ is
in a cycle of period 1 for half its possible initial conditions
and in a cycle of period 2 for the other half.
Thus, for large $N$ there will on average
be many isolated spins in cycles of length two or four.
It follows that it will be overwhelmingly likely that the
cycle length of
the entire system will be divisible by two.

Our $N$'s are not very large.
But $N=10$ is large enough so that isolated spins are
likely, and they play a substantial role in making our
cycle-lengths mostly divisible by several factors of two.
As figure~\ref{Osci:ratio} demonstrates, when $K$ is small,
the ratio of even to odd cycles increases rapidly as $N$ is
increased, consistent with the idea that local
structures have a substantial effect at small $K$.

Factors of three are only slightly harder to obtain.
Imagine, for example, a spin coupled
only to itself via
$$F_i({\Sigma}) = \pm \sigma_i ~.$$
One initial condition for
such a spin results in a cycle length of one, while the other
three result in a cycle length of three.
These and other structures can then give us our observed factors of
three.

\begin{figure}
\centerline{\epsfxsize=3.5in
\epsfbox{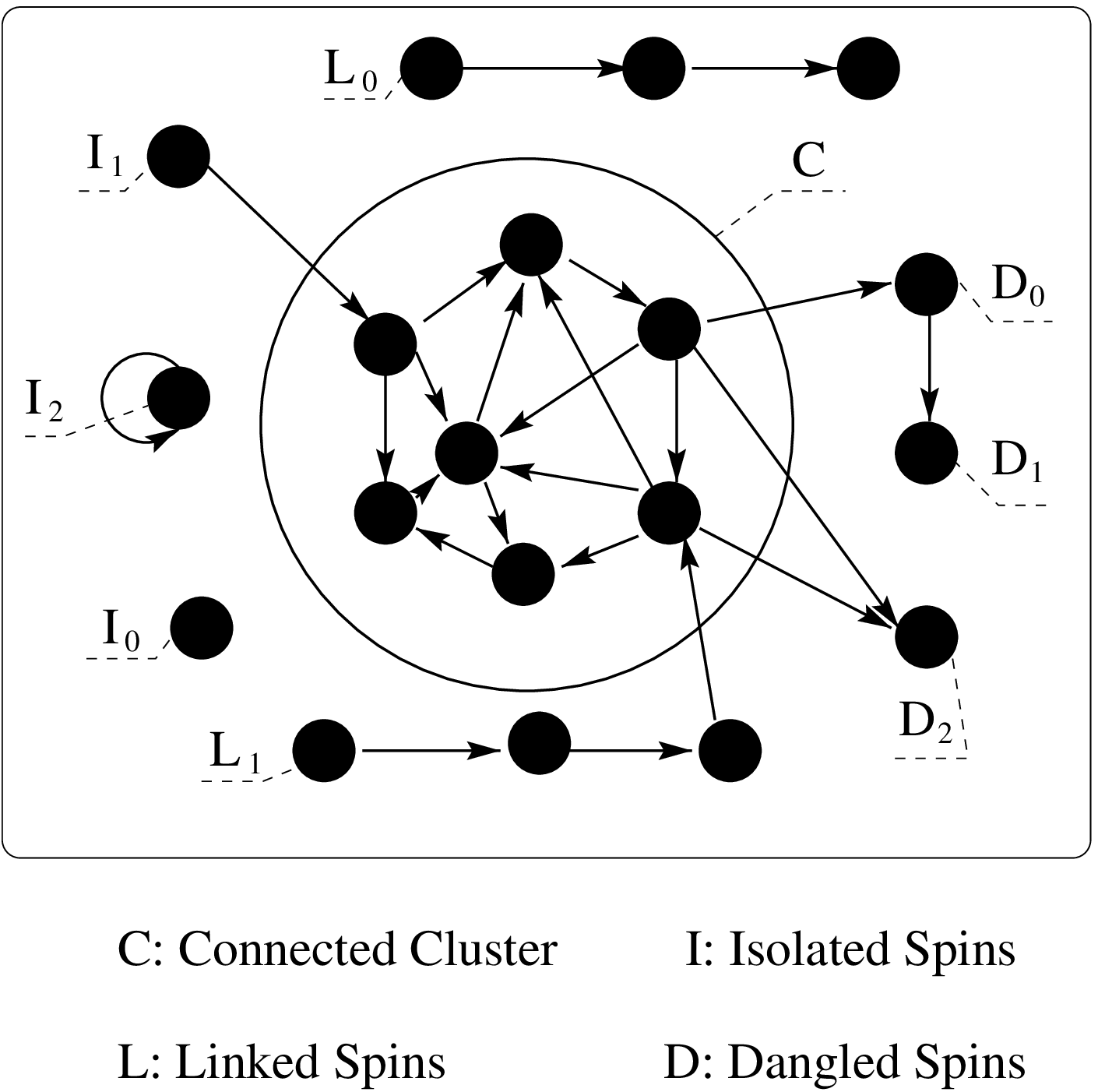}}
\caption{This figure shows the simplest kinds of clusters.  The arrows
show the flow of information. The cluster inside the circle
labeled `C' is connected. That means it contains many spins intimately
linked by complex interactions. The 
clusters labeled `$I_0$', `$I_1$' and `$I_2$' are single spins, isolated
from the influence of any others.
As we have depicted them, the `$I_0$' conveys no information to the connected
cluster while `$I_1$' 
brings information from the isolated cluster to the connected cluster. We
also depict the linkages `$L_0$' and `$L_1$', slightly more complex structures
which also undergo their own independent motion. The $D$'s indicate
dangling structures. }
\label{clusters}
\vskip 0.75cm
\end{figure}

Other isolated structures can
be formed that have a substantial influence on system properties.
For example, one can have a
structure like the
linkage `$L_0$' shown in Fig. \ref{clusters}.  A linkage is a sequence of
spins linked
together by coupling of the form $\pm \sigma$.  The arrow indicates that the
value of the
spin on at the tail changes the values of the spin at the head.  A linkage
like `$L_0$' has
no effect on the cluster.
One can also have a linkage like `$L_1$' which directly
influences the cluster.  We shall discuss the effect of linkages in more
detail in the
next chapter.  For now we merely point out that these, and other, more complex
structures have a simple effect upon the overall periods of the system.
Each small
isolated structure has its own short period.  The motion for the entire
system must be a multiple of the periods for each of the isolated
structures.  Since the small
structures all tend to have periods two and three and four, it is no wonder
that these
factors appear very often in the observed periods of Fig. \ref{fin41grf}.

\subsubsection{Dangling structures}
It is also possible to have isolated structures which are influenced by the
connected cluster
but do not influence it themselves.  For example, consider a spin,
$\sigma_i$, whose input function
$ F_i({\Sigma})$ depends only on spins that are not causally related to
$\sigma_i$. Then the extra spin `dangling on' the cluster produces a
simple effect upon the periods
of the entire system. To deal with this situation, we prove a theorem.

\begin{theo}
\label{theo1}
Consider two sequences $F^t$ and $\sigma^t$. We assume that both of them
are sequences of $+1$'s and $-1$'s, that $F^t$ is periodic with period
$T_F$, and that
the $\sigma$'s depend upon the $F$'s according to
\beq
\sigma^{t+1}=F^t\cdot\sigma^{t-1}.
\label{dy}
\eeq
Then $\sigma$ is periodic with period $T_\sigma$, where
$T_\sigma$ can only be $T_F$, $2T_F$, or $4T_F$.

We prove this by showing that (A) $T_\sigma$ divides
$4T_F$ evenly, and (B) $T_F$ divides $T_\sigma$ evenly.
To prove (A), note that
\beq
\sigma^{t+2T_F}=\Pi(t) ~ \sigma^t~,
\eeq
where
\beq
%\Pi(t) = \Pi_{m=1}^{T_F}F^{t+2T_F-2m+1}~.
\Pi(t) = F^{t+2T_F-1}F^{t+2T_F-3}\ldots F^{t+3}F^{t+1}~.
\eeq
When $T_F$ is odd, $\Pi(t)$ is a product over a full period of the
$F$'s. Therefore,
it is independent of time.  If it is +1, the sequence of
$\sigma$'s repeats after $2 T_F$.
If it is -1, then $\sigma^t$ repeats after $4 T_F$.
If $T_F$ is even, then
$$
\Pi(t)=\left [ \Pi_{m=1}^{T_F/2}F^{t+T_F-2m+1} \right ]^2=1~.
$$
Therefore, when $T_F$ is even,
$\sigma^{t+2T_F}=\sigma^{t}$.
Thus, for both even and odd $T_F$,
$\sigma$ is periodic with a period $T_\sigma$
that divides $4T_F$ evenly.

Now we prove (B), that $T_F$ divides $T_\sigma$ evenly.
For any  $t$, we have
\beq
\sigma^{T_\sigma+t}&=&\Pi'(t)~\sigma^{t}  \text{\quad with}
\nonumber \\
\Pi'(t)&=& F^{T_\sigma+t-1}\cdots F^{t+3} F^{t+1}~.  \nonumber
 \eeq
The periodicity of $\sigma$ implies that $ \Pi'(t) =1 $ for all
$t$. Note that, for all $t$,
\beq
\Pi'(t+2) =F^{T_\sigma+t+1}~ F^{t+1} ~ \Pi'(t)~. \nonumber
\eeq
Thus, $F^{T_\sigma+t}=F^{t}$ for all $t$, and
$T_F$ must divide $T_\sigma$ evenly.

Q.E.D.

\end{theo}

Theorem~\ref{theo1} enables us to see additional mechanisms for the
multiplication of cycle
lengths by factors of two and four.  Look at the structures dangling from
the end of a connected cluster as also shown in Fig. \ref{clusters}.  
A dangling cluster consisting of
a single spin can multiply the periods produced
by the connected
cluster by a factor of one, two, or four. The two-spin dangle can, at
most, produce a
lengthening by a factor of eight.  Thus, these dangles produce
additional mechanisms for
the appearance of factors of two in the period of the entire system.

\section{Some local structures}
\label{sec:k=1}
In the preceding two sections we saw that small $K$ reversible Boolean
nets exhibit large oscillations, both in the behavior of the
Hamming distance as a function of time and the orbit period distribution
as a function of period $l$.  We explained the latter as an effect of small
localized structures.  In this section, we discuss the localized structures
in more detail.  Our purpose is both to explain the oscillations
exhibited above, as well as to explore some of
the structures' statistical properties.

\subsection{The structures and their coupling}
\label{locals}
\begin{figure}
\centerline{\epsfxsize=3.5in\epsfbox{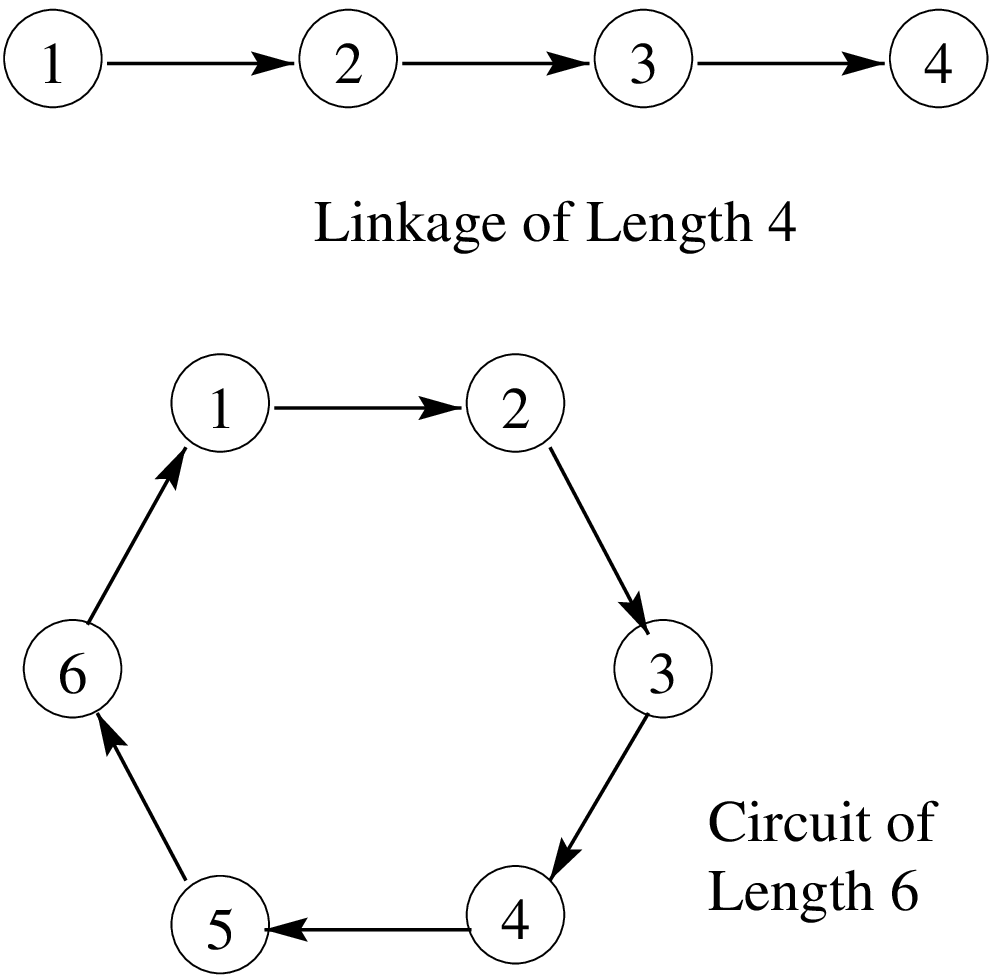}}
\caption{ Two kinds of basic  structures.  Each circle
represents a spin; the spin at the end of an arrow depends on the
spin from which the arrow starts. }
\label{local}
\vskip 0.75cm
\end{figure}

Two kinds of elements, depicted in figure~\ref{local}, serve as a basis for
a description of the behavior at $K=1$. These same elements, called
\mbox{\em linkages} and \mbox{\em circuits}, may also be
expected to dominate the behavior at small $K$.
Others
have discussed the effect of structures like these for the dissipative
Kauffman model\cite{BP98,FL88,FK88,AB00}.
The behavior is, of course, different in the reversible case.

\subsubsection{The structures}
\begin{figure}
\centerline{\epsfxsize=3.5in\epsfbox{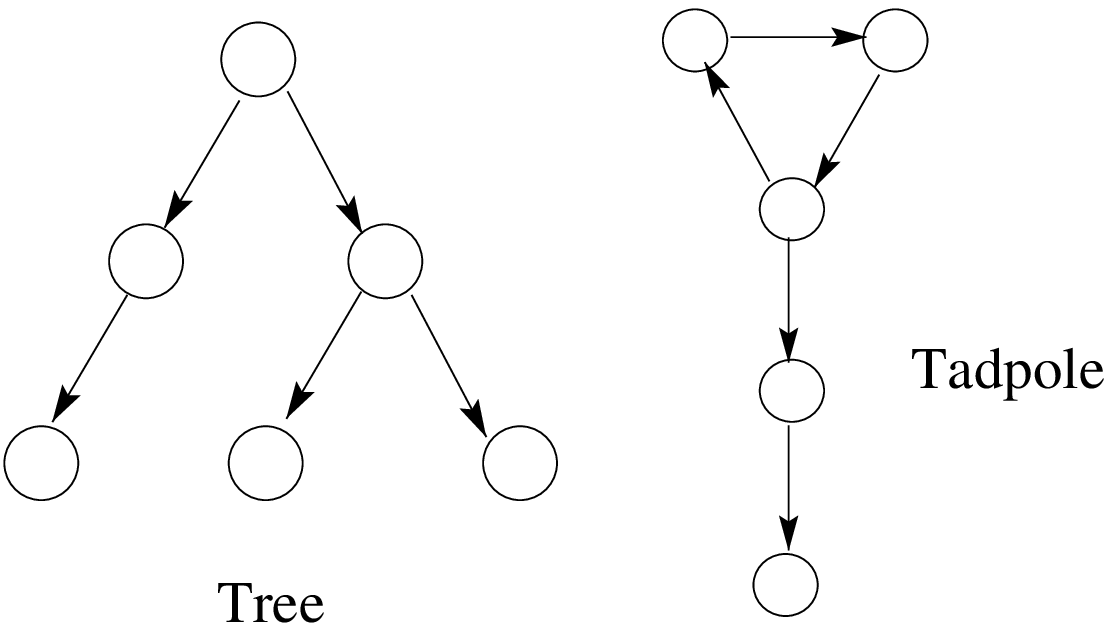}}
\caption{Possible ways in which the basic structures may be tied together.
Both the tree and the tadpole are possible structures at K=1.  However, if
the arrows were reversed, these structures would only become possible at
higher $K$. }
\label{neglected}
\vskip 0.75cm
\end{figure}

A linkage is  $L$ spins, $\sigma_{j},~j=1,\ldots,L$,
coupled together in a linear array.
The first spin has an input function, $F_1$, that is either $+1$ or
$-1$.  Information is transferred from one spin to the next, for all $j$
bigger than one, with a coupling function
\begin{equation}
F_j(\Sigma)=\pm \sigma_{j-1} \qquad \mbox{for }\qquad j=2, 3,
\ldots, L~.
\label{LinkStructure}
\end{equation}
A circuit of length $L$ is like a linkage except that the last spin is
coupled to the first with
\begin{equation}
F_1(\Sigma)=\pm \sigma_L.
\end{equation}
These basic
structures may also be tied together, as for example, in the tree and
the tadpole shown in figure \ref{neglected}.
In this section, we shall analyze
the basic structures.  We shall, however, not explore the effects of the
multiplicity of structures which can be present at one time, nor of the
different ways these structures can be coupled.

\subsubsection{Local structures at $K=1$.}

For $K=1$, since two of the four Boolean functions of one
variable do not depend on the input, half of all spins
have dynamics that do not depend on the state of any other spin.
Therefore, typical local structures in networks with $K=1$
tend to contain only a few spins.

For a given $K$, the numbers of linkages and circuits depend only on the
placement of the couplings and on the function choices,
and  do not depend on whether the dynamical
equation is dissipative or reversible. Flyvbjerg and Kjaer~\cite{FK88} have
calculated the number of circuits of length $L$
exactly in the context of the dissipative Kauffman model.
However, they did not calculate the number of linkages, doubtless because
linkages of all lengths have trivial dynamics in the dissipative Kauffman
model.

The number of spins
in linkages is of order $N$,
while the number of spins in circuits is of order unity.
This scaling follows readily via the following argument.
When $K=1$ each spin has one input chosen from $N$ possibilities,
and the spin depends on the input only if its function is
$\pm\sigma$, which occurs with probability $1/2$.
The probability that a spin depends on any other spin is of order one, while
the probability that it depends upon itself is of order $1/N$. Consequently,
there are of order $N$ spins linked to other spins,  but only
of order one spin linked to itself.  An extension of this argument implies
that there are of order $N$ linkages, but only of order one circuit in the
entire system.
Tree-like structures can be formed from
linkages alone, and tadpoles from combinations of linkages and circuits.
The former
have a behavior which is qualitatively similar to a single linkage,
the latter have properties which are closer to that of circuits.

\subsection{Theorems for the analysis of linkages and circuits.}
\label{rev:k=1}

There are three theorems which greatly facilitate our work with linkages and
circuits.

We have already stated one of these as our theorem \ref{theo1} in
Sec.~\ref{sec:Oscillations}.
This theorem tells us how the period of a base structure is affected by a spin
dangling on the end.
The period of the base is multiplied by one or two
if the base period is even and one, two, or four if it is odd. By
direct calculation, one sees that a linkage of length one will have a
period of one or two or four.  Hence, all linkages have
periods that are integer powers of two.

The other two theorems apply more specifically to linkages and
circuits.  They are:
\begin{theo}[Equivalent Structure Theorem]
For a linkage or a circuit,
switching the function assignment $F_{j}(\sigma)$
from $\sigma$ to $-\sigma$ does not affect the dynamics of the
structure in the sense that
we can always find a one-to-one map from the
state-space of the local structure to itself, relating a cycle before
switching the function assignment to a cycle with the same cycle length
after the switch.

Proof: For each $F_{j}(\sigma)$ which has the value $-\sigma$,
change the variable $\sigma_{j-1}$ into $-\sigma_{j-1}$.  Q.E.D.
\end{theo}
Note: This change leaves all Hamming distances and cycle lengths unchanged.
Hence, for the examination of these quantities we need only consider the
case in which the linkage functions are $F_{j}(\Sigma)= \sigma_{j-1}$
and not the cases in which  $F_{j}(\Sigma)=- \sigma_{j-1}$.

Theorem~\ref{superposition} is a superposition principle
that applies to systems with $K=1$.
\begin{theo}[Superposition Principle]\label{superposition}
Consider a
network with $K=1$.
The equations of motion are of the form
\begin{equation}
\sigma_j^{t+1}=F_j(\Sigma^t)\sigma_j^{t-1}~,
\label{eqofmotion}
\end{equation}
Let
\begin{eqnarray}
f_j &=& 1 \text{~~if~~} F_j(\Sigma^t) = +1 \text{~or~} \sigma_i^t
\text{\quad for some~} i(j)~ \nonumber \\
f_j &=& -1 \text{~~if~~} F_j(\Sigma^t) = -1 \text{~or~}
-\sigma_i^t
\text{\quad for some~} i(j)~.
\label{bond}
\end{eqnarray}
Let  $\sigma_j^t$ and $\mu_j^t$ both be solutions to the
equations of motion for the same realization of the $K=1$ couplings. Then
\begin{equation}
 \rho_j^t= f_j ~ \sigma_j^t \mu_j^t
\end{equation}
is also a solution.

Proof:  This multiplication rule can be immediately verified on each bond in
equation~\ref{eqofmotion}. Therefore it must be true globally.
\end{theo}
Note that the change of variables from $\sigma$ to $(\ln f~ \sigma)/(i \pi) $
converts this multiplicative superposition into the  kind of
additive superposition principle which arises for all linear equations.

%\texttt{Condition~\ref{bond} is not as restricting as it seems to be. Note
%that in the three representing structures, only minus linkages does not
%satisfy the condition. However, we may reconstruct the evolution of a minus
%linkage of length $L$ using a plus linkage of length $L+1$ with
%$\sigma_1^0=\sigma_1^1=-1$, since $F_2(\sigma_1)\equiv-1$. Therefore,
%although the theorem only applies to standard circuits and plus
%linkages, we can still superpose different solutions for
%plus linkages and standard circuits to draw information about
%cycle-length distributions and Hamming distance behaviors for all
%linkages and curcuits. This applicability of superposition is based on the
%equivalent structure theorem and the argument above concerning minus
%linkages.}

\subsection{Hamming distance for linkages and circuits.}
\label{hamk=1}

One can make considerable progress in the Hamming distance problem for
long linkages and circuits. The Equivalent Structure Theorem implies
that for each $L$ we need to study only one circuit problem and two
linkage problems. The latter differ by
their first coupling, which can take on the two values $F_1=\pm 1$.

First we show that the Hamming distance can be obtained by solving a
single linkage or a circuit problem. Consider a linkage or circuit with
a solution $\Sigma^t = \{\sigma_j^t\}$, and let $\mathcal{M}^t=\{\mu_j^t\}$ be another solution with initial
conditions that are identical to those for $\Sigma^t$ at $t=0$
and have, at $t=1$, a single spin different, the spin
with index $J$.  Then defining
\begin{equation}
  \rho_j^t= \sigma_j^t \mu_j^t ~,
\end{equation}
all the values of $\rho_j^0$ and $\rho_j^1$ are unity
except for $\rho_J^1=-1$.   The variable $\rho_j^t$ is minus one
whenever the two paths differ, so the Hamming distance
between $\mathcal{M}$ and $\Sigma$ at time $t$ is
\begin{equation}
D(t)= \sum_{j=1}^L \half(1-\rho_j^t)~.
\end{equation}
The subsequent
history of $\rho$ is calculated starting from the  initial
data and the equations of motion
\begin{equation}
\rho_j^{t+1}=  \rho_{j-1}^{t}  \rho_j^{t-1}
  \text{\quad for \quad} i=2,3, \dots, L~.
\label{90}
\end{equation}
For the linkage, all the $\rho_j^t$ are one if $j<J$ and $(-1)^t$ if $j=J$.
The picture just stops short at $i=L$; nothing is meaningful to the right of
this value of $i$. For the circuit, we apply the periodic boundary conditions
$ \rho_{j+L}^t = \rho_j^t$.

\begin{figure}
\centerline{\epsfxsize=3.5in\epsfbox{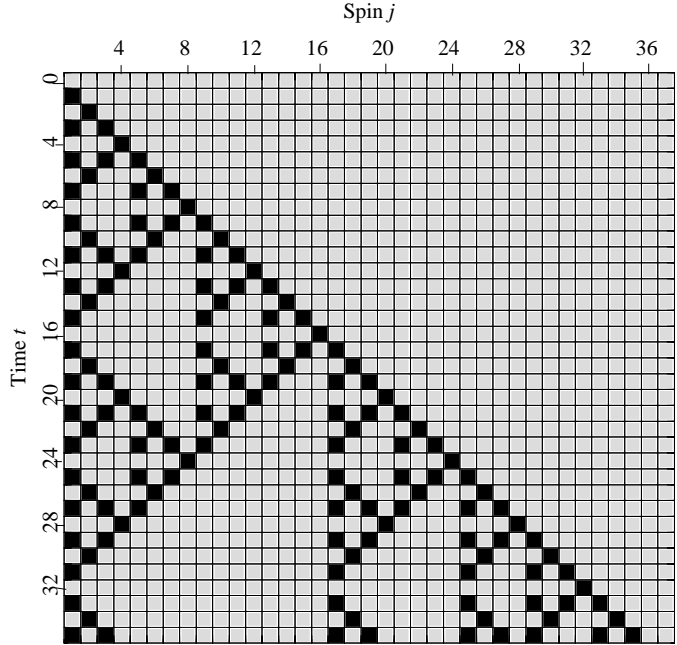}}
\caption{Cascading Effect of a Single Spin Flip. Here, $\rho_j^t$ is
plotted for a situation in which the flip occurs in a very large linkage or
circuit.  The grey squares denote $\rho=1$; the black, $\rho=-1$.}
\label{sir}
\vskip 0.75cm
\end{figure}

The result for a very large linkage or circuit is shown in Figure~\ref{sir}.
This figure shows the familiar Sierpinski
gasket~\cite{mandelbrot77,mandelbrot83},
rotated  ninety degrees from the conventional representation.
In fact, equation
\ref{90} is equivalent to rule
$90$ in Wolfram's cellular automaton scheme, which is known to generate
a Sierpinski gasket~\cite{wolfram83}.

\begin{figure}
\centerline{\epsfxsize=3.5in\epsfbox{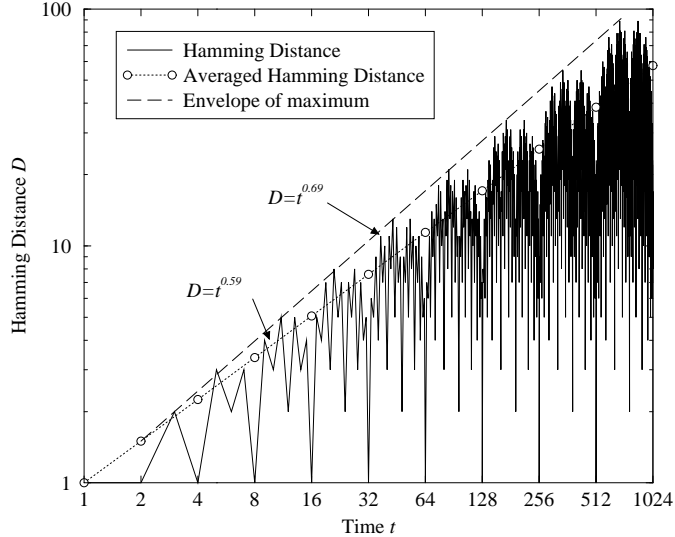}}
\caption{Numerically calculated Hamming distance versus time for a linkage with
Sierpinski gasket dynamics (solid line). The average of the Hamming
distance in each binary decade grows as $t^{\left(\ln_2 3\right)}\approx
t^{0.59}$ (dotted line), while the maximum value of the Hamming distance
in each binary decade grows as $t^{\log_2\frac{1+\sqrt{5}}{2}} \approx
t^{0.69}$ (dashed line).}
\label{hammS}
\vskip 0.75cm
\end{figure}

Figure
\ref{hammS} shows a plot of Hamming distance $D(t)$ versus iteration
number $t$ for the gasket.
The distance is unity at times $2^n$ and
its average in the binary decade $[2^n, 2^{n+1}]$ is
$\left(\frac{3}{2}\right)^{n-1}$. Thus, this average 
increases as $t^{(\log_2 3-1)}\approx t^{0.59}$~\cite{Wol0}. 
The maximum value of
$D(t)$ in the range $[2^{n-1}, 2^{n}]$ is the Fibonacci number $F_n$,
where $F_1=1$, $F_2=2$, and $F_j=F_{j-1}+F_{j-2}$, which scales as
$t^{\log_2\left(\frac{1+\sqrt{5}}{2}\right)}\approx t^{0.69}$. Both the
average of $D(t)$ and the maximum value of $D(t)$ in each
binary decade can be obtained analytically by exploiting the fractality
and symmetry properties of the Sierpinski gasket.  

To calculate the Hamming distance in a linkage of finite length $L$,
started by flipping a spin $J$ sites into the
linkage, note that no information travels to sites $\sigma_i$ with
$i<J$, and that the dynamics are cut off at the last site $i=L$. 
If the distance from $J$ to the 
right hand boundary is between $2^n$
and $2^{n+1}$, the motion repeats with a period $2^{n+2}$.
We shall see below how this information may be converted into
statements about the distribution of periods for the linkage.

The Hamming distance behavior for the circuits can be understood by
seeing the circuit picture as one in which the gasket picture is folded
onto itself an infinite number of times,
with the folding distance being, $L$, the
size of the circuit.  Because two superposed black dots cancel and make
a gray one, the folding can produce
complex patterns.  Nonetheless, we know that the Hamming
distance is unity at times which are of the form $2^n$, for
integer $n$, since at these times there is but one black dot.
Furthermore, the typical or average Hamming distance must grow no more
rapidly than $t^{0.59}$,
since this growth is bounded above by the number of black dots in the
non-overlapping gasket.
We shall come back to this fold-over picture in
order to assess periods of circuit motion.

\begin{table}
\begin{center}
\begin{tabular}{|c|c||c|c|c|c|c|c|}
\hline
  &    &
\multicolumn{5}{c|} {number of cycles with period}
\\ \cline{3-7}
L & $F_1$ & $l=$1 & $l=$2 & $l=$4 & $l=$8 & $l=$16 \\ \hline \hline
%1 & -1 &   &   & 1 &   &     &   \\
%\hline
1 &  1 & 2 & 1 &   &   &        \\
\hline
%2 & -1 &   &   &   & 2 &     &   \\
%\hline
2 & 1  & 2 & 1 & 3 &   &        \\
\hline
%3 & -1 &   &   &   & 8 &     &   \\
3 &  1 & 2 & 1 & 3 & 6 &        \\
\hline
%4 & -1 &   &   &   &   & 16  &   \\
4 &  1 & 2 & 1 & 3 & 30&        \\
\hline
%5 & -1 &   &   &   &   & 64  &   \\
5 &  1 & 2 & 1 & 3 & 30& 48     \\
\hline
%6 & -1 &   &   &   &   & 256 &   \\
6 & 1  & 2 & 1 & 3 & 30& 240    \\
\hline
%7 & -1 &   &   &   &   & 1024&   \\
7 & 1  & 2 & 1 & 3 & 30& 1008   \\
\hline
%8 & -1 &   &   &   &   &     &2048\\
8 &  1 & 2 & 1 & 3 & 30& 4080    \\
\hline
\end{tabular}
\end{center}
\caption{Numbers of orbits of different periods
for linkages of lengths $L$ from $1$ to $8$ with $F_1=+1$.}
\label{tab:k1:l}
\vskip 0.75cm
\end{table}

%table2.tex
\begin{table}
\begin{center}
\begin{tabular}{|c|l|}
\hline
$L$ & Cycle Length $l$'s (Multiplicities)\\
\hline
1 & 1(1) 3(1) \\
\hline
2 & 1(1) 3(1) 6(2)\\
\hline
3 &
1(1) 3(1) 5(3) 15(3)\\
\hline
4 &
1(1) 3(1) 6(2) 12(20)\\
\hline
5 &
1(1) 3(1) 17(15) 51(15)\\
\hline
6 &
1(1) 3(1) 5(3) 6(2) 10(24) 15(3) 30(126)\\
\hline
7 &
1(1) 3(1) 7(9) 9(28) 21(9) 63(252)\\
\hline
8 &
1(1) 3(1) 6(2) 12(20) 24(2720)\\
\hline
\end{tabular}
\end{center}
\caption{Complete enumeration of all possible orbits for circuits
of lengths $L \le 8$.
The numbers in the body of the table are the
possible cycle lengths and, in parentheses,
the number of different cycles of
that length.
%For a circuit of length $l$, there are also $2^{2l}$ different
%states of the system, so that the sum of the products of cycle lengths
%and the number of different cycles of the length must equal $2^{2l}$.
}
\label{tab:k1:c}
\vskip 0.75cm
\end{table}

\subsection{Cycle length distributions for linkages.}

Table~\ref{tab:k1:l} shows the distribution of cycle-lengths for the first
few linkages. As expected, the linkages all have periods which are powers of
two.  We now calculate the distribution of cycle-lengths.

\begin{figure}
\centerline{\epsfxsize=4.5in\epsfbox{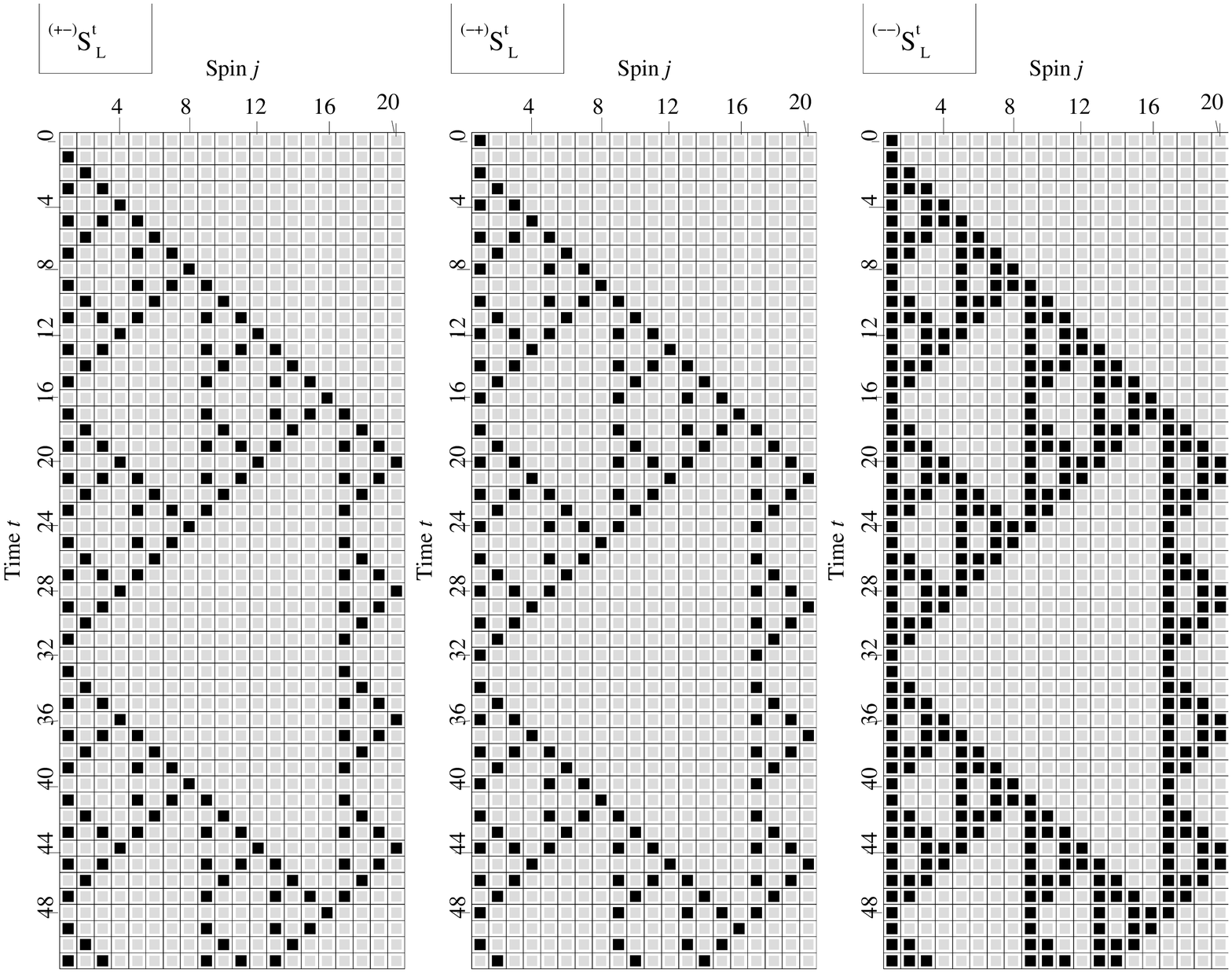}}
\caption{Primitive sequences ${^{(+-)}\mathcal{S}_L^t}$,
${^{(-+)}\mathcal{S}_L^t}$, and ${^{(++)}\mathcal{S}_L^t}$ for a linkage
of length $L=20$.}
\label{sirLp}
\vskip 0.75cm
\end{figure}

First we show that the orbit period of a linkage
is determined entirely by the root spin, defined to be the
spin $\sigma_J$ with the smallest value of $J$
that has an initial condition other than
$(+1,+1)$.
It is useful to define the $(-+)$ \emph{primitive sequence}
${^{(-+)}\mathcal{S}_L^t}$ for a linkage of length $L$
to be the values $\{\sigma_i^t\}$ for $i=1,....L$,
$t=2,3,....$ that one gets starting a  linkage of length $L$ with $F_1=1$
from the initial
condition with all $\sigma_i^0=\sigma_i^1=1$ for all $i$ except for
$\sigma_1^0=-1$.
The {\em infinite primitive sequence}
$^{(-+)}\mathcal{S}^t$ is the $L\rightarrow\infty$
limit of the ${^{(-+)}\mathcal{S}_L^t}$.
The $(+-)$ primitive sequence is obtained by starting
a  linkage with $F_1=1$ with the initial
conditions $\sigma_i^0$ and $\sigma_i^1=1$
except for $\sigma_1^1=-1$.
For all $t>1$, the $(+-)$ primitive sequence is obtained
from the $(-+)$ primitive sequence by changing the
origin of time by one.
The primitive sequence $^{(--)}\mathcal{S}^t$
can be obtained by superposing
$^{(+-)}\mathcal{S}^t$ and $^{(-+)}\mathcal{S}^t$. The primitive sequences
${^{(-+)}\mathcal{S}_L^t}$,
${^{(-+)}\mathcal{S}_L^t}$ and ${^{(++)}\mathcal{S}_L^t}$
for a linkage of length
$L=20$ are shown in figure~\ref{sirLp}.
We use $\mathcal{S}^t$ to denote any one of these three
primitive sequences.

As demonstrated in fig.~\ref{sirLp},
the $l^{th}$ spin ($l>1)$ in the $(-+)$ primitive sequence
cycles with the period
\begin{equation}
p(l)= 2^{[ \log_2 (l-1) ] + 2}~,
\label{littlep}
\end{equation}
where $[x]$ denotes the greatest integer less than or equal to x.
Obviously the period of the $l^{th}$ spin
in the $(+-)$ primitive sequence is also $p(l)$, and
it is straightforward to verify that in the $(--)$
primitive sequence the $l^{th}$ spin cycles with the
period $p(l)$ if $l>1$.

Now we show that changing the initial conditions
of spins other than the root spin does not change the
period $p(l)$.
We do this by noting that the time development from
any initial condition can be written
as the sum of primitive sequences with different spatial origins.
The form of $p(l)$ guarantees that any superposed structure
has a period that is either equal to $p(l)$ or
else divides $p(l)$ evenly.
Since the time development of the $j^{th}$ spin
can be written as a superposition
$\sigma_j^t = A^t B^t$, where
$A^{t+p(l)}=A^t$ and $B^{t+p(l)}=B^t$, clearly
$T_\sigma$, the period of the $\sigma^t$,
must either be equal
to $p(l)$ or else divide it evenly.

However, we still need to show that $T_\sigma$
is not smaller than $p(l)$ (that this is not trivial
can be seen by noting that superposing the two
period-four sequences $(++--++--\ldots)$
and $(--++--++\ldots)$ yields a period-one
sequence).
First note that the period cannot decrease if $T_B$, the
period of the $B^t$, is less than $T_A$,
the period of the $A^t$.
This follows because if $T_B<T_A$,
then $B^{t+T_A/2}=B^{t}$, and
$\sigma_j^{t+T_A/2} = A^{t+T_A/2}B^{t+T_A/2}
= A^{t+T_A/2}B^t \ne A^tB^{t}$.
Therefore, we need only consider the case in which
the periods $T_A$ and $T_B$ are equal.

We use the Sierpinski gasket form of the
primitive sequences to show that
superposing sequences $A$ and $B$ does
not decrease the period when $l>2$.
The spins in a primitive sequence
which have equal periods have time evolutions that
map out triangular shapes in the $i-t$ plane.
For example, in the $(+-)$ primitive sequence
the values $\sigma_i^t$ in which
$i=2^n+j$ and $t=3\cdot2^n-j$ with $0< j\le 2^n$
for any integer $n$ have $\sigma_i^t=-1$.
Superposing a sequence with a larger value of
$i$ in the range $[2^n+1,2^{2n}]$
cannot eliminate the $-1$ for the
smallest value of $i$ in this range.
Since $\sigma_i^{t}=1$ for all $t$ with
$3\cdot2^n+2\le t\le 5\cdot2^n$ when
$i$ obeys $2^n< i\le 2^{2n}$,
combining any of these sequences cannot
reduce the period.
A similar argument applies to the $(-+)$
primitive sequence using the lower edge of
triangles in the gasket.

Thus we have shown that if a linkage has $F_1=1$ and the
root spin has an initial condition other than $(+1,+1)$,
the cycle length of the $l^{th}$ spin with $l>2$ is
$p(l) = 2^{[\log_2(l-1)+2]}$.
Since the spin $\sigma_2^t$ in the
${(--)}$ primitive sequence has the time evolution
$(+1,+1,-1,-1,\ldots)$, which is the same time
development of the root spin $\sigma_1^t$ in a linkage
with $F_1=-1$, it follows immediately that the
cycle length of the $l^{th}$ spin in a linkage with
$F_1=-1$ is 
\begin{equation}
P(l)=2^{[\log_2(l)]+2}~.
\end{equation}

Now we are in a position to find the distribution
of orbit lengths for different linkages.
Let $N_{F_1=-1}(L,l)$ and $N_{F_1=1}(L,l)$
be the number of orbits
of length $l$ in a linkage of length $L$ with
input functions $-1$ and $+1$, respectively.
Calculating $N_{F_1=-1}(L,l)$ is simple.
We know that all the orbits in this linkage have
length $P(L)$, and that there are $2^{2L}$ points
in the phase space altogether, so the orbit length
distribution is
\begin{equation}
N_{F_1=-1}(L,l) =
\delta_{l,P(L)} \frac{2^{2L}}{P(L)}~.
\end{equation}

To obtain the distribution of cycle lengths for
a linkage of length $L$ with $F_1=1$,
we must consider initial conditions
in which $\sigma_1^0=\sigma_1^1=+1$, where the
behavior is identical to that of a shorter linkage.
The three other initial conditions for
$\sigma_1$ yield the
orbit period $p(L)$, so
$N_{F_1=1}(L,l)$ satisfies the recursion
relation
\begin{equation}
 N_{F_1=1}(L,l) = 3\frac{(2^{2L-2})}{p(L)}\delta_{l,p(L)}
+N_{F_1=1}(L-1,l)~.
\end{equation}
The solution to this recursion relation has
$N_{F_1=1}(1,l)=2\delta_{l,1}+\delta_{l,2}$,
$N_{F_1=1}(2,l)=2\delta_{l,1}+\delta_{l,2}+3\delta_{l,4}$,
and, for $L>2$,
\begin{eqnarray}
 N_{F_1=1}(L,l) =
\frac{4^{(2^{[\log_2(L-1)]}-1 )}}
{2^{[\log_2(L-1)]}}
\left ( 4^{L-2^{[\log_2(L-1)]}}-1\right )\delta_{l,p(L)}
\nonumber \\
+
2\delta_{l,1} + \delta_{l,2}+
\sum_{n=0}^{[\log_2(L-1)]-1}
\frac{4^{2^n-1}}{2^n}(4^{2^n}-1)\delta_{l,2^{n+2}}~.
\end{eqnarray}
This result agrees with our simulational data shown in Table
\ref{tab:k1:l}.

\begin{figure}
\centerline{\epsfxsize=3.5in
\epsfbox{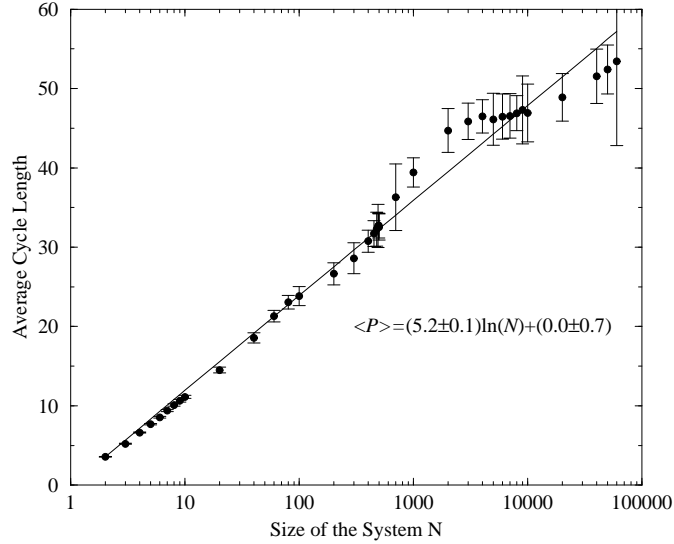}}
\caption{Average cycle length $\expect{P}$ plotted against the size of the
system $N$ for the reversible model with $K=1$.
The average is over 100 randomly chosen realizations,
with 100 cycles randomly generated for each realization. The solid line
is a fit to the numerical data assuming a logarithmic dependence:
$\expect{P}=(5.2\pm 0.1)\mbox{ln}N+(0.0\pm0.7)$.}
\label{finavek1}
\vskip 0.75cm
\end{figure}

We can connect these results to the way the typical cycle length
scales with system size $N$ for $K=1$.
Since the realization average of the number of circuits per realization
does not change with $N$, we expect the linkages to dominate the scaling,
with the circuits only
giving a constant multiplicative factor in the typical cycle length.
For $K=1$, the probability of finding a
linkage of length $L$ decreases exponentially as $L$
increases. Therefore, $L_{max}(N)$, the longest linkage in a system of
$N$ spins, obeys
\begin{equation}
L_{max}(N)\propto \log(N)~.
\label{l_max}
\end{equation}
We just argued that the typical cycle length of
a linkage of length $L$ is
\begin{equation}
P(L)\propto 2^{[\log_2{L}]}~.
\label{P(l)equation}
\end{equation}
Eqs.~(\ref{l_max}) and~(\ref{P(l)equation}) together
imply that the typical length of the cycles for $K=1$
should increase logarithmically with $N$.
Our simulation results, shown in
figure~\ref{finavek1}, are consistent with this scaling.

\subsection{Dynamics of circuits}
Table~\ref{tab:k1:c} gives the distribution of cycle lengths for circuits
for the lowest
values  of $L$.

\begin{figure}
\centerline{\epsfxsize=3.5in\epsfbox{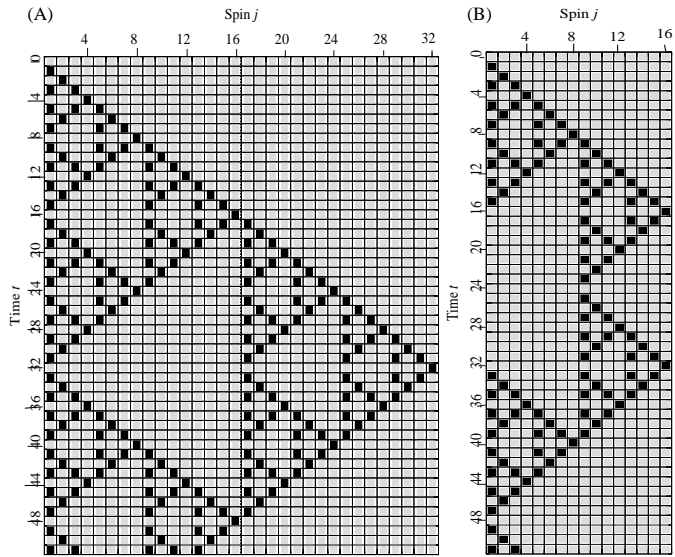}}
\caption{Dynamical development of deviations in a linkage (A) and 
in a circuit (B).  }
\label{sirL}
\vskip 0.75cm
\end{figure}

It is hard to find the general form of distribution of cycle lengths for
this situation.\footnote{Algebraic techniques introduced in
reference~\cite{Wol1} may be of use to solve the problem.} In the
special case in which $L$ is of the form $2^n$, one can indeed see what
will happen.  For example,
consider the structure shown in figure
\ref{sir} if $n=4$. Recall that the circuit structure makes the picture
wrap around at $j=16, 32, \cdots$.  After $32$ steps,
a structure arises between
$j=32$ and
$j=48$ which can cancel out the similar structure sitting between $j=0$ and
$j=16$.  Then, after 48 steps, another structure
beyond $j=48$ unfolds itself and repeats the structure seen for small
times.  Hence the circuit has a
closed cycle. The resulting picture is shown in Figure~\ref{sirL}(B).
Something very similar happens after 63 steps for $L$=7, but
we have not fully
unraveled the form of the resulting behavior for general $L$.

The result shown in Table \ref{tab:k1:c} partially answers a question which
arose in our discussion
of oscillations in the number of cycles for $K=2$.  In that case, we found
that the preponderance of
cycles had periods which were divisible by three.  The problem is that
neither linkages nor the trees
put together from linkages have cycle lengths that are divisible by three.
Large connected
clusters might be expected to have a smoothly varying distribution of cycle
lengths and hence only
one third of these would be divisible by three.  But small circuits can
partially save the day. The
table shows that the factor of three indeed arises quite often in these
periods.  We are pleased to
see this, but not totally satisfied.  There are only a few circuits
(typically one) in a system
with $N=10$ and $K=2$.  Somewhere there must be an additional source of
factors of three.

\section{Discussion}
\label{discussion}

In this paper we have studied the behavior of the Hamming distance
and of the cycle length distribution in Boolean nets with
time-reversible dynamics.
We obtain strong evidence for the presence of a phase transition
at a critical value $K_c \simeq 1.65$ and present analytic bounds on
$K_c$.
We observe large oscillations in the behavior of the Hamming distance
as a function of time and of the cycle length distribution as a function
of cycle length when $K$ is relatively small.
We propose that local structures, or small groups of spins with dynamics
that do not depend on any other spins in the systems, play a crucial
role in giving rise to these fluctuations.
By analyzing local structures, significant analytic insight
can be obtained regarding both the
oscillatory behavior of the average number of cycle lengths averaged
over realizations as well as the minima in the Hamming distance that
occur at times that are multiples of
$2^n$ for integer $n$.

Although we feel that we have a qualitative understanding of the
phenomena exhibited by this model, some unexplored issues remain.
It would be desirable to develop a
more quantitative understanding of when
period-three cycles are favored.
Obtaining sharper bounds on the
critical value $K_c$ would also be desirable.
 
We thank Raissa D'Souza for useful conversations.
SNC and LPK gratefully acknowledge financial support from the
National Science Foundation.
SNC thanks the Aspen Center for Physics
for hospitality during the preparation of this manuscript.

\appendix

\end{document}